\documentclass{article}
\usepackage{spconf}
\usepackage{amssymb}
\usepackage{graphicx,subfigure}
\usepackage[centertags]{amsmath}
\usepackage{epsfig,algorithm,amsmath,amssymb,color,subfigure,times,url}
\ninept

\newtheorem{theorem}{Theorem}
\newtheorem{lemma}{Lemma}

\newtheorem{definition}{Definition}
\long\def\comment#1{}

\begin{document}
\renewcommand{\textfraction}{0}

\title{Analyzing Least Squares and Kalman Filtered Compressed Sensing
} 
\name{Namrata Vaswani
\thanks{This research was partially supported by NSF grant ECCS-0725849}
}
\address{Dept. of ECE, Iowa State University, Ames, IA, namrata@iastate.edu}
\date{}
\maketitle \thispagestyle{empty}


\setlength{\arraycolsep}{0.03cm}
\newcommand{\xhat}{\hat{x}}
\newcommand{\xpred}{\hat{x}_{t|t-1}}
\newcommand{\Ppred}{P_{t|t-1}}
\newcommand{\ty}{\tilde{y}_t}
\newcommand{\tty}{\tilde{y}_{t,\text{res}}}
\newcommand{\tw}{\tilde{w}_t}
\newcommand{\ttw}{\tilde{w}_{t,f}}
\newcommand{\betahat}{\hat{\beta}}

\newcommand{\ypast}{y_{1:t-1}}
\newcommand{\sone}{S_{*}}
\newcommand{\sinf}{{S_{**}}}
\newcommand{\smax}{S_{\max}}
\newcommand{\smin}{S_{\min}}
\newcommand{\samax}{S_{a,\max}}
\newcommand{\Nhat}{{\hat{N}}}

\newcommand{\Dnum}{D_{num}}
\newcommand{\pss}{p^{**,i}}
\newcommand{\fr}{f_{r}^i}

\newcommand{\A}{{\cal A}}
\newcommand{\Z}{{\cal Z}}
\newcommand{\B}{{\cal B}}
\newcommand{\R}{{\cal R}}
\newcommand{\reg}{{\cal G}}
\newcommand{\const}{\mbox{const}}

\newcommand{\trace}{\mbox{tr}}

\newcommand{\hsim}{{\hspace{0.0cm} \sim  \hspace{0.0cm}}}
\newcommand{\he}{{\hspace{0.0cm} =  \hspace{0.0cm}}}

\newcommand{\vect}[2]{\left[\begin{array}{cccccc}
     #1 \\
     #2
   \end{array}
  \right]
  }

\newcommand{\matr}[2]{ \left[\begin{array}{cc}
     #1 \\
     #2
   \end{array}
  \right]
  }
\newcommand{\vc}[2]{\left[\begin{array}{c}
     #1 \\
     #2
   \end{array}
  \right]
  }

\newcommand{\gdot}{\dot{g}}
\newcommand{\Cdot}{\dot{C}}
\newcommand{\re}{\mathbb{R}}
\newcommand{\n}{{\cal N}}  
\newcommand{\N}{{\overrightarrow{\bf N}}}  
\newcommand{\chat}{\tilde{C}_t}
\newcommand{\chati}{\chat^i}

\newcommand{\cmin}{C^*_{min}}
\newcommand{\twi}{\tilde{w}_t^{(i)}}
\newcommand{\twj}{\tilde{w}_t^{(j)}}
\newcommand{\wi}{{w}_t^{(i)}}
\newcommand{\twio}{\tilde{w}_{t-1}^{(i)}}

\newcommand{\tWi}{\tilde{W}_n^{(m)}}
\newcommand{\tWj}{\tilde{W}_n^{(k)}}
\newcommand{\Wi}{{W}_n^{(m)}}
\newcommand{\tWio}{\tilde{W}_{n-1}^{(m)}}

\newcommand{\ds}{\displaystyle}

\newcommand{\SAR}{S$\!$A$\!$R }
\newcommand{\MAR}{MAR}
\newcommand{\MMRF}{MMRF}
\newcommand{\AR}{A$\!$R }
\newcommand{\GMRF}{G$\!$M$\!$R$\!$F }
\newcommand{\DTM}{D$\!$T$\!$M }
\newcommand{\MSE}{M$\!$S$\!$E }
\newcommand{\RCS}{R$\!$C$\!$S }
\newcommand{\uomega}{\underline{\omega}}
\newcommand{\y}{v}
\newcommand{\x}{w}
\newcommand{\lu}{\mu}
\newcommand{\g}{g}
\newcommand{\s}{{\bf s}}
\newcommand{\bft}{{\bf t}}
\newcommand{\refmap}{{\cal R}}
\newcommand{\totrefl}{{\cal E}}
\newcommand{\beq}{\begin{equation}}
\newcommand{\eeq}{\end{equation}}
\newcommand{\bdm}{\begin{displaymath}}
\newcommand{\edm}{\end{displaymath}}
\newcommand{\hatz}{\hat{z}}
\newcommand{\hatu}{\hat{u}}
\newcommand{\tilz}{\tilde{z}}
\newcommand{\tilu}{\tilde{u}}
\newcommand{\hhatz}{\hat{\hat{z}}}
\newcommand{\hhatu}{\hat{\hat{u}}}
\newcommand{\tilc}{\tilde{C}}
\newcommand{\hatc}{\hat{C}}
\newcommand{\tim}{n}

\newcommand{\ssp}{\renewcommand{\baselinestretch}{1.0}}
\newcommand{\defd}{\mbox{$\stackrel{\mbox{$\triangle$}}{=}$}}
\newcommand{\goes}{\rightarrow}
\newcommand{\tends}{\rightarrow}
\newcommand{\defn}{\triangleq} 
\newcommand{\se}{&=&}
\newcommand{\sdefn}{& \defn  &}
\newcommand{\sle}{& \le &}
\newcommand{\sge}{& \ge &}
\newcommand{\plusminus}{\stackrel{+}{-}}
\newcommand{\Ey}{E_{Y_{1:t}}}
\newcommand{\ey}{E_{Y_{1:t}}}

\newcommand{\equivto}{\mbox{~~~which is equivalent to~~~}}
\newcommand{\nonzero}{i:\pi^n(x^{(i)})>0}
\newcommand{\nonzeroc}{i:c(x^{(i)})>0}

\newcommand{\supn}{\sup_{\phi:||\phi||_\infty \le 1}}
\newtheorem{remark}{Remark}
\newtheorem{example}{Example}
\newtheorem{ass}{Assumption}
\newtheorem{proposition}{Proposition}

\newtheorem{fact}{Fact}
\newtheorem{heuristic}{Heuristic}
\newcommand{\eps}{\epsilon}
\newcommand{\bd}{\begin{definition}}
\newcommand{\ed}{\end{definition}}
\newcommand{\udq}{\underline{D_Q}}
\newcommand{\td}{\tilde{D}}
\newcommand{\epsinv}{\epsilon_{inv}}
\newcommand{\al}{\mathcal{A}}

\newcommand{\bfx} {\bf X}
\newcommand{\bfy} {\bf Y}
\newcommand{\bfz} {\bf Z}
\newcommand{\ddas}{\mbox{${d_1}^2({\bf X})$}}
\newcommand{\ddbs}{\mbox{${d_2}^2({\bfx})$}}
\newcommand{\dda}{\mbox{$d_1(\bfx)$}}
\newcommand{\ddb}{\mbox{$d_2(\bfx)$}}
\newcommand{\xinc}{{\bfx} \in \mbox{$C_1$}}
\newcommand{\eqa}{\stackrel{(a)}{=}}
\newcommand{\eqb}{\stackrel{(b)}{=}}
\newcommand{\eqe}{\stackrel{(e)}{=}}
\newcommand{\leqc}{\stackrel{(c)}{\le}}
\newcommand{\leqd}{\stackrel{(d)}{\le}}

\newcommand{\leqa}{\stackrel{(a)}{\le}}
\newcommand{\leqb}{\stackrel{(b)}{\le}}
\newcommand{\leqe}{\stackrel{(e)}{\le}}
\newcommand{\leqf}{\stackrel{(f)}{\le}}
\newcommand{\leqg}{\stackrel{(g)}{\le}}
\newcommand{\leqh}{\stackrel{(h)}{\le}}
\newcommand{\leqi}{\stackrel{(i)}{\le}}
\newcommand{\leqj}{\stackrel{(j)}{\le}}

\newcommand{\w}{{W^{LDA}}}
\newcommand{\halpha}{\hat{\alpha}}
\newcommand{\hsigma}{\hat{\sigma}}
\newcommand{\slmax}{\sqrt{\lambda_{max}}}
\newcommand{\slmin}{\sqrt{\lambda_{min}}}
\newcommand{\lmax}{\lambda_{max}}
\newcommand{\lmin}{\lambda_{min}}

\newcommand{\da} {\frac{\alpha}{\sigma}}
\newcommand{\chka} {\frac{\check{\alpha}}{\check{\sigma}}}
\newcommand{\sumo}{\sum _{\underline{\omega} \in \Omega}}
\newcommand{\distance}{d\{(\hatz _x, \hatz _y),(\tilz _x, \tilz _y)\}}
\newcommand{\col}{{\rm col}}
\newcommand{\rcs}{\sigma_0}
\newcommand{\CalR}{{\cal R}}
\newcommand{\df}{{\delta p}}
\newcommand{\dq}{{\delta q}}
\newcommand{\dZ}{{\delta Z}}
\newcommand{\pprime}{{\prime\prime}}

\newcommand{\vn}{N}

\newcommand{\bv}{\begin{vugraph}}
\newcommand{\ev}{\end{vugraph}}
\newcommand{\bi}{\begin{itemize}}
\newcommand{\ei}{\end{itemize}}
\newcommand{\ben}{\begin{enumerate}}
\newcommand{\een}{\end{enumerate}}
\newcommand{\be}{\protect\[}
\newcommand{\ee}{\protect\]}
\newcommand{\bean}{\begin{eqnarray*} }
\newcommand{\eean}{\end{eqnarray*} }
\newcommand{\bea}{\begin{eqnarray} }
\newcommand{\eea}{\end{eqnarray} }
\newcommand{\nn}{\nonumber}
\newcommand{\ba}{\begin{array} }
\newcommand{\ea}{\end{array} }
\newcommand{\ep}{\mbox{\boldmath $\epsilon$}}
\newcommand{\epp}{\mbox{\boldmath $\epsilon '$}}
\newcommand{\Lep}{\mbox{\LARGE $\epsilon_2$}}
\newcommand{\und}{\underline}
\newcommand{\pdif}[2]{\frac{\partial #1}{\partial #2}}
\newcommand{\odif}[2]{\frac{d #1}{d #2}}
\newcommand{\dt}[1]{\pdif{#1}{t}}
\newcommand{\urho}{\underline{\rho}}

\newcommand{\spc}{{\cal S}}
\newcommand{\tspc}{{\cal TS}}

\newcommand{\uv}{\underline{v}}
\newcommand{\us}{\underline{s}}
\newcommand{\uc}{\underline{c}}
\newcommand{\utheta}{\underline{\theta}^*}
\newcommand{\ualpha}{\underline{\alpha^*}}

\newcommand{\uxy}{\underline{x}^*}
\newcommand{\uxyj}{[x^{*}_j,y^{*}_j]}
\newcommand{\arcl}[1]{arclen(#1)}
\newcommand{\one}{{\mathbf{1}}}

\newcommand{\uxyjt}{\uxy_{j,t}}
\newcommand{\E}{\mathbb{E}}

\newcommand{\rhomat}{\left[\begin{array}{c}
                        \rho_3 \ \rho_4 \\
                        \rho_5 \ \rho_6
                        \end{array}
                   \right]}
\newcommand{\deltat}{\tau} 
\newcommand{\deltatt}{\Delta t_1}
\newcommand{\ceil}[1]{\ulcorner #1 \urcorner}

\newcommand{\xxi}{x^{(i)}}
\newcommand{\txi}{\tilde{x}^{(i)}}
\newcommand{\txj}{\tilde{x}^{(j)}}

\newcommand{\mi}[1]{{#1}^{(m,i)}}

\setlength{\arraycolsep}{0.05cm}
\newcommand{\Section}[1]{  \vspace{-0.1in} \section{#1}  \vspace{-0.07in} } 
\newcommand{\Subsection}[1]{  \vspace{-0.1in} \subsection{#1}  \vspace{-0.05in} } 
\newcommand{\Subsubsection}[1]{  \vspace{-0.1in} \subsubsection{#1}  \vspace{-0.05in} } 

\begin{abstract} 
In recent work, we studied the problem of causally reconstructing time sequences of spatially sparse signals, with unknown and slow time-varying sparsity patterns, from a limited number of linear ``incoherent" measurements. We proposed a solution called Kalman Filtered Compressed Sensing (KF-CS). The key idea is to run a reduced order KF only for the current signal's estimated nonzero coefficients' set, while performing CS on the Kalman filtering error to estimate new additions, if any, to the set. KF may be replaced by Least Squares (LS) estimation and we call the resulting algorithm LS-CS.
In this work, (a) we bound the error in performing CS on the LS error and (b) we obtain the conditions under which the KF-CS (or LS-CS) estimate converges to that of a genie-aided KF (or LS), i.e. the KF (or LS) which knows the true nonzero sets.
\end{abstract}

\begin{figure}[t!]
\centerline{
\subfigure{
\label{sims_nodel}
\epsfig{file = 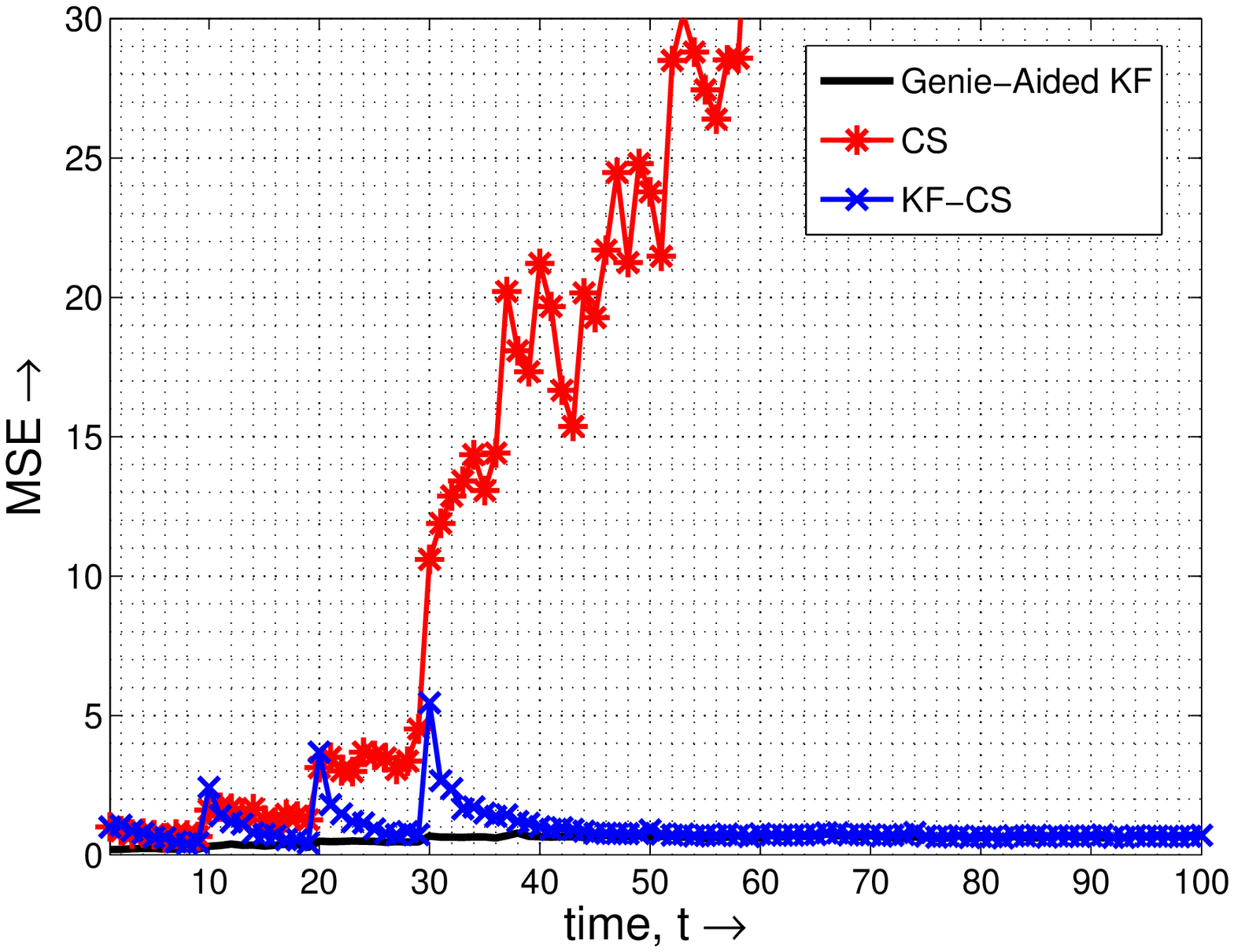, height = 3cm,width=4.25cm}
}
\subfigure{
\label{sims_del}
\epsfig{file = 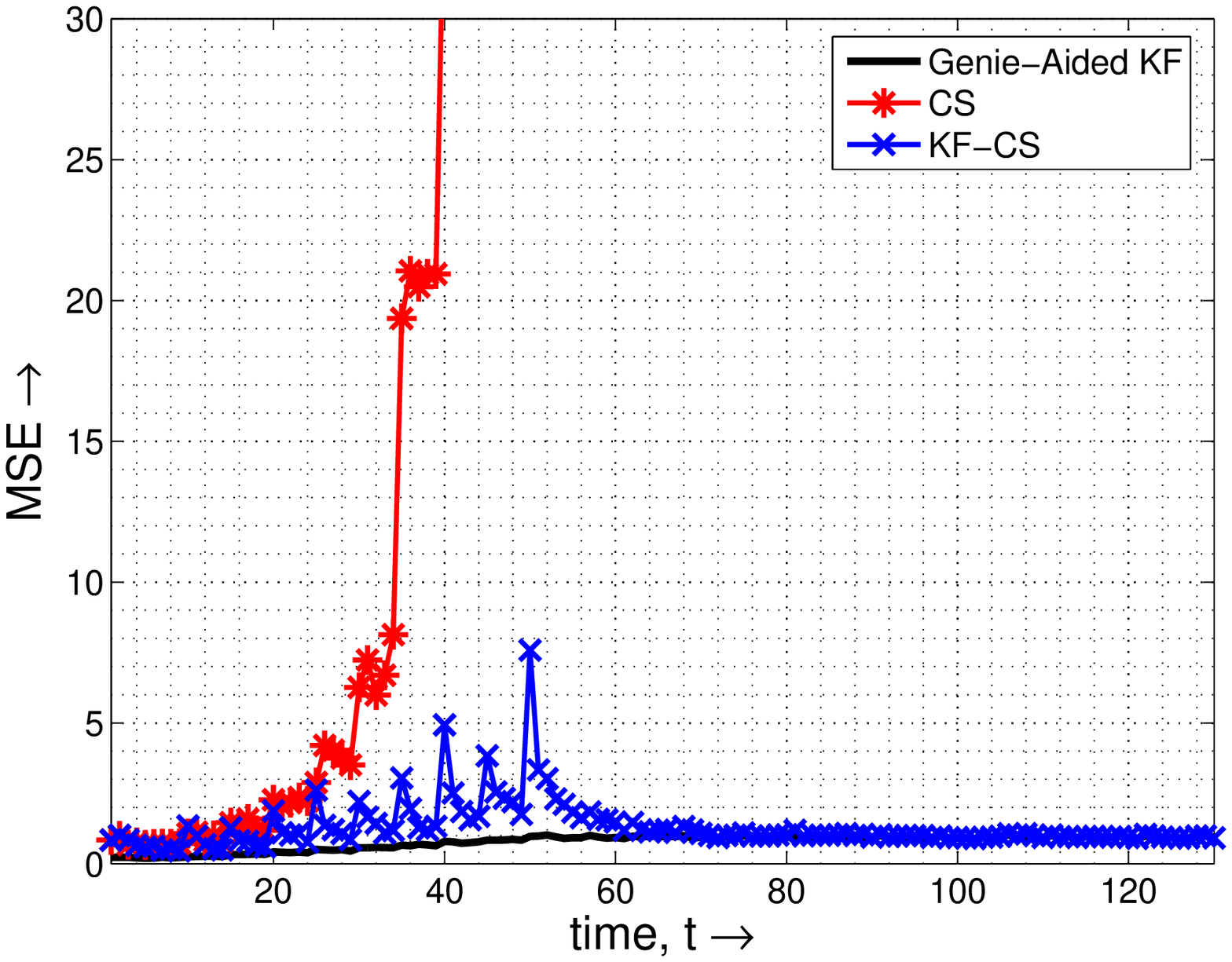, height = 3cm,width=4.25cm} 
}
}
\vspace{-0.15in}
\caption{\small{MSE plots comparison. Fig. \ref{sims_nodel}: Large KF-CS error occurs at and after the new addition times, $t=10, 20, 30$. But once the addition is detected, the error gradually reduces to that of GA-KF (or slightly higher). The error of simple CS (labeled as CS) is much larger (max value 45). Fig. \ref{sims_del}: Simple CS error beyond $t=50$ when $S_t \ge 26$ than is much larger in the Fig. \ref{sims_nodel} (max value 425).
}}
\vspace{-0.195in}
\label{fig}
\end{figure}

\Section{Introduction}
In recent work \cite{kfcsicip}, we studied the problem of causally reconstructing time sequences of sparse signals, with unknown and slow time-varying sparsity patterns, from a limited number of noise-corrupted ``incoherent" measurements. We proposed a solution called Kalman Filtered Compressed Sensing (KF-CS). With the exception of CS \cite{dantzig} and of \cite{rozell}, most other work \cite{sparsedynamicMRI,singlepixel} treats the entire time sequence of signals/images as a single spatiotemporal signal and performs CS to reconstruct it. This is a non-causal solution and also has high computational cost. On the other hand, if the number of observations is small, performing CS \cite{dantzig} at each time (simple CS) incurs much larger error than KF-CS, see Fig. \ref{fig}.
Potential applications of KF-CS include making dynamic MRI real-time (causal and fast enough) \cite{sparsedynamicMRI,kfcsmri}; real-time video imaging using a single-pixel camera \cite{singlepixel}; or real-time tracking of temperature, or other, time-varying fields using sensor networks that transmit random projections of the field \cite{nowaknoise}.

In this work, in Sec. \ref{kfcsdefine}, we describe a simple modification of KF-CS \cite{kfcsicip} and introduce its non-Bayesian version, Least Squares (LS)-CS. Our {\em key contributions} are: (a) in Sec. \ref{compare}, we bound the error in performing CS on the LS error in the observation and compare it with that for performing CS on the observation (simple CS), and (b) in Sec. \ref{converge}, we obtain the conditions under which the KF-CS (or LS-CS) estimate converges to that of a genie-aided KF (or LS). Simulation comparisons are given in Sec. \ref{sims}.%

{\em Problem Definition. }
The problem definition is the same as in \cite{kfcsicip}. 
Let $(z_t)_{m \times 1}$ denote the spatial signal of interest at time $t$ and $(y_t)_{n \times 1}$, with $n<m$, denote its noise-corrupted observation vector at $t$. The signal, $z_t$, is sparse in a given sparsity basis (e.g. wavelet) with orthonormal basis matrix, $\Phi_{m \times m}$, i.e. $x_t \defn \Phi' z_t$ is a sparse vector (only $S_t << m$ elements of $x_t$ are non-zero). The observations are ``incoherent" w.r.t. the sparsity basis of the signal, i.e.
\bea
y_t = A x_t + w_t, \ A \defn H \Phi, \ \E[w_t]=0, \ \E[w_t w_t']= \sigma_{obs}^2 I 
\label{obsmod}
\eea
and all $S_t$-column sub-matrices of $A$ ``approximately orthonormal", i.e. $\delta_{S_t} < 1$ \cite[eq. (1.3)]{dantzig}. $w_t$ is independent of $x_t$ and is i.i.d, $\forall t$.

Let $N_t$ denote the the support set of $x_t$, i.e. the set of its non-zero coordinates and let $T_t \defn \hat{N}_t$ denote its estimate. Also, let $\Delta_t$ denote the undetected nonzero set at time $t$, i.e.  $\Delta_t \defn N_t \setminus T_{t-1}$ and let $\hat{\Delta}_t$ denote its estimate. Thus $T_t = T_{t-1} \cup \hat{\Delta}_t$. Let $S_t \defn |N_t|$ where $|.|$ denotes the size of a set.
Also, for any set $T$, let $(v)_T$ denote the $|T|$ length sub-vector containing the elements of $v$ corresponding to the indices in the set $T$. For a matrix $A$, $A_T$ denotes the sub-matrix obtained by extracting the columns of $A$ corresponding to the indices in $T$. We use the notation $(Q)_{T_1,T_2}$ to denote the sub-matrix of $Q$ containing rows and columns corresponding to the entries in $T_1$ and $T_2$ respectively. 
$T^c$ denotes the complement of $T$ w.r.t. $[1:m]$, i.e. $T^c \defn [1:m] \setminus T$. $\phi$ refers to the empty set. $'$ denotes transpose.
{\em The $m \times m$ matrix $I_T$ is defined as follows: $(I_T)_{T,T} = I$ where $I$ is a $|T|$-identity matrix while  $(I_T)_{T^c,[1:m]} = \mathbf{0}$,  $(I_T)_{[1:m], T^c} = \mathbf{0}$.}

The nonzero coefficients' set $N_t$ changes slowly over time. For the currently nonzero coefficients of $x_t$, $(x_t)_{N_t}$, we assume a spatially i.i.d. Gaussian random walk model, while the rest of the coefficients remain constant, i.e.
\bea
x_0 = \mathbf{0}, \ \ x_t \  \se   \ x_{t-1} + \nu_t, \ \ \nu_t \sim \n(0,Q_t),   \ \ Q_t = \sigma_{sys}^2 I_{N_t} \ \ \ \
\label{sysmod}
\eea
where $\nu_t$ is temporally i.i.d.. {\em The current nonzero set, $N_t$, is unknown $\forall t$}.
Our goal is to recursively get the best estimates of $N_t$ and $x_t$ (or equivalently of the signal, $z_t = \Phi x_t$) using $y_1, \dots y_t$.

 \begin{algorithm*}[t!]
\caption{ Kalman Filtered Compressive Sensing (KF-CS)}
{\bf Initialization:} Set $\xhat_{0} = 0$, ${P}_{0}=0$, $T_0= \phi$ (if unknown) or equal to the known support. For $t > 0$, do,
\ben

\item {\bf Temporary Kalman prediction and update. } Implement (\ref{kftmp}) using $\hat{Q}_t = \sigma_{sys}^2 I_{T_{t-1}}$.
\label{firstkf}

\item {\bf Compute Additions using CS. } Compute the KF error, $\tty \defn y_t - A \xhat_{t,tmp}$. Check if $FEN \defn \tty' \Sigma_{fe,t}^{-1} \tty > \alpha_{fe}$. If it is,
\ben
\item {\em Do CS on $\tty$ followed by thresholding}, i.e. compute $\hat\Delta_t$ using (\ref{ds}), (\ref{dsthresh}).
 The new estimated support is $T_{t} = T_{t-1} \cup \hat\Delta_t$.




\een
\label{csfestep}

\item  {\bf Kalman prediction and update. } Implement (\ref{kfbig}) using $\hat{Q}_t = \sigma_{sys}^2 I_{T_{t}}$.
\label{secondkf}
\ben
\item (KF-CS with final LS): If $T_{t} \neq T_{t-1}$, implement (\ref{kfbig}) using $\hat{Q}_t = \infty I_{T_{t}}$, i.e. set $\xhat_t = (A_{T_t}'A_{T_t})^{-1} A_{T_t}'y_t$ and $(P_t)_{T_t,T_t} = (A_{T_t}'A_{T_t})^{-1} \sigma_{obs}^2$,   $(P_t)_{T_t^c,:}=0$,  $(P_t)_{:,T_t^c} = 0$.
\label{secondls}
\een

\item {\bf Compute Deletions. } If $T_{t} == T_{t-1} \dots == T_{t-k}$ (nonzero set has not changed for long enough, i.e. w.h.p. KF stabilized),%
\ben
\item Check for ``zero" coefficients, i.e. compute $\hat\Delta_Z = \{i \in T_{t} : \sum_{\tau=t-k'+1} ^t (\xhat_{\tau,i})^2 / k' < \alpha_z \}$ with $k' < k$.
Set $T_{t} \leftarrow T_{t} \setminus \hat\Delta_Z$. Set $(\xhat_t)_{\hat\Delta_Z} = 0$. Set $(P_{t})_{\hat\Delta_Z,[1:m]} = 0$ and $(P_{t})_{[1:m],\hat\Delta_Z} = 0$. 

\een

\item {\bf Output $T_t$, $\xhat_t$ and the signal estimate, $\hat{z}_t = \Phi \xhat_t$.} Increment $t$ and go to the first step.
\een
\label{kfcs}
\end{algorithm*}

\Section{Kalman Filtered CS and Least Squares CS}
\label{kfcsdefine}
We describe a simple modification of KF-CS \cite{kfcsicip} and introduce Least Squares CS. Let $\xhat_{t|t-1}, \xhat_{t}$, $K_t$ and $P_{t|t-1},P_t$ denote the predicted and updated state estimates at time $t$, the Kalman gain and the  prediction and updated error covariances given by the KF in KF-CS (since KF-CS does not always use the correct value of $Q_t$, $P_{t|t-1}$ or  $P_t$ are not equal to the actual covariances of $x_t - \xpred$ or $x_t - \xhat_t$).%

\Subsection{Modified Kalman Filtered Compressed Sensing (KF-CS)}
\label{kfcsmod}
KF-CS can be summarized as running a KF for the system in (\ref{obsmod}), (\ref{sysmod}) but with $Q_t$ replaced by $\hat{Q}_t = \sigma_{sys}^2 I_{T_t}$. The new additions, if any, are estimated by performing CS on the Kalman filtering error, $\tty$.

At time $t$, we first run a ``temporary" Kalman prediction and update step using $\hat{Q}_t = \sigma_{sys}^2 I_{T_{t-1}}$, i.e. we compute
\bea
K_{t,tmp} \se (P_{t-1} + \hat{Q}_t) A' (A (P_{t-1} + \hat{Q}_t) A' + \sigma_{obs}^2 I)^{-1}  \nn \\
 \xhat_{t,tmp} \se (I - K_{t,tmp} A) \xhat_{t-1} + K_{t,tmp} \ y_t 
 \     
\label{kftmp}
\eea
Let $T \defn T_{t-1}$. The filtering error is
\bea
\tty \defn y_t - A \xhat_{t,tmp} = A_{\Delta_t} (x_t)_{\Delta_t} + A_{T} (x_t - \hat{x}_t)_T + w_t
\eea
As explained in \cite{kfcsicip}, if the filtering error norm is large, there is a need to estimate $\Delta_t$.
One can rewrite $\tty$ as  $\tty = A \beta_t + w_t$, where $\beta_t \defn [(x_t - \xhat_t)_T, (x_t)_{\Delta_t}, 0_{ (T \cup {\Delta_t})^c }]$ is a ``sparse-compressible" signal with a ``large" or ``non-compressible" nonzero part, $(x_t)_{\Delta_t}$, and a ``small" or ``compressible" nonzero  part, $(x_t - \xhat_t)_T$. The Dantzig selector (DS) \cite{dantzig} followed by thresholding can be applied to detect the ``non-compressible" nonzero part as follows: 
\bea
\label{ds}
\betahat_t \se \arg \min_\beta ||\beta||_1, s.t. \ ||A'(\tty - A \beta) ||_\infty \le \lambda_m \sigma_{obs}  \\
\label{dsthresh}
\hat{\Delta}_t \se \{i \in T_{t-1}^c: \betahat_{t,i}^2 > \alpha_a \}
\eea
where $\lambda_m \defn \sqrt{2 \log m}$ and $\alpha_a$ is the addition threshold.
Thus, the estimated support set at time $t$ is $T_{t} = T \cup \hat\Delta_t = T_{t-1} \cup \hat\Delta_t$.

Next we run the Kalman prediction/update using $\hat{Q}_t = \sigma_{sys}^2 I_{T_t}$: 
\bea
P_{t|t-1} \se P_{t-1} + \hat{Q}_t, \ \
K_t = P_{t|t-1} A' (A P_{t|t-1} A' + \sigma_{obs}^2 I)^{-1} \nn \\
P_t \se (I - K_t A) P_{t|t-1}  \nn \\
\xhat_t \se (I - K_t A) \xhat_{t-1} + K_t y_t
\label{kfbig}
\eea
with initialization $P_0 = \mathbf{0}_{[1:m],[1:m]}, \ \ \xhat_0 = \mathbf{0}_{[1:m]}$.
\begin{remark}
For easy notation, in (\ref{kftmp}),(\ref{kfbig}) we write the KF equations for the entire $x_t$. But actually we are running a reduced order KF for only the coefficients in $T$ ($T \equiv T_{t-1}$ for (\ref{kftmp}) and $T \equiv T_{t}$ for (\ref{kfbig}).
\end{remark}



\Subsubsection{Deleting Zero Coefficients}
\label{kfcswithdel}
If the addition threshold, $\alpha_a$, is not large enough, occasionally there will be some false additions (coefficients whose true value is zero but they wrongly get added due to error in the CS step). Also, there may be coefficients that actually become and remain zero.
All such coefficients need to be detected and removed from $T_t$ to prevent unnecessary increase in $|T_t|$. Increased $|T_t|$ implies smaller minimum eigenvalue of $A_{T_t}'A_{T_t}$ and thus increased estimation error. The increase is especially large if $A_{T_t}'A_{T_t}$ is close to becoming singular.

 One possible way to detect if a coefficient, $i$, is zero is to check if the magnitude of its estimates in the last few time instants is small, e.g. one can check if $\sum_{\tau=t-k'+1}^t (\xhat_{\tau,i})^2 / k' < \alpha_z$. This scheme would be fairly accurate (small enough false alarm and miss probabilities), if the estimation error, $e_{\tau,i} = x_{\tau,i} - \xhat_{\tau,i}$ is small enough, for all $\tau \in [t-k'+1,t]$. If we check for zeroing only when $T_t$ has not changed for long enough (w.h.p. this implies that all past additions have been detected, i.e. $T_t=N_t$, and the KF for $T_t$ has stabilized), the variance of $e_{\tau,i}$ would be approximately equal to $(P_{t})_{i,i} < {\sigma_{obs}^2}/{\lambda_{min}(A_T'A_T)}$, i.e. it would be small enough.

When a coefficient, $i$, is detected as being zero, we remove it from $T_t$, we set $\xhat_{t,i} = 0$ and we set $(P_{t})_{i,[1:m]} = 0$,  $(P_{t})_{[1:m],i} = 0$. We summarize the entire KF-CS algorithm in Algorithm \ref{kfcs}.

\Subsection{Least Squares CS: Non-Bayesian KF-CS}
\label{lscs}
In applications where training data is not be available to learn the prior model parameters required by KF-CS, one can use a non-Bayesian version of KF-CS i.e. replace the KF in KF-CS by Least Squares (LS) estimation. The LS step is also faster than the KF step.




\Section{Analyzing CS on LS Error (LSE)}
\label{compare}

Let $T \defn T_{t-1}$ and $\Delta \defn \Delta_t = N_t \setminus T_{t-1}$.
The true nonzero sets at any time, $N_t$, are assumed to be non-random. But $T = T_{t-1}$ is a random variable since its value depends on $y_{t-1}$ and $T_{t-2}$ (or equivalently on $y_{1:t-1}$). We use $\E[ \cdot ]$ to denote expectation w.r.t. all random quantities ($y_{1:t}, x_{1:t}$ at time $t$) while using $\E[ \cdot | y_{1:t-1}]$ to denote the expected value conditioned on $y_{1:t-1}$. Conditioned on $y_{1:t-1}$, the set $T$, and hence also the set $\Delta = N_t \setminus T$, is known.

The key difference between simple CS and LS-CS is that simple CS applies (\ref{ds}) on $y_t = A x_t + w_t$ to estimate the $|N_t|$-sparse signal, $x_t$, while LS-CS applies (\ref{ds}) on the LS error (LSE), $\tty := y_t - A \xhat_{t,tmp} = A \beta_t + w_t$ to estimate $\beta_t := x_t -  \xhat_{t,tmp}$, where $\xhat_{t,tmp} = (A_T'A_T)^{-1}A_T'y_t$.  $\beta_t = [(x_t - \xhat_{t,tmp})_T , (x_t)_{\Delta} , 0_{{T \cup \Delta}^c}] = (A_T'A_T)^{-1}A_T'(A_\Delta (x_t)_\Delta + w_t) , (x_t)_{\Delta} , 0_{{T \cup \Delta}^c}]$ is what we call a ``sparse-compressible" signal: it is $|T \cup \Delta|$-sparse but, if the sparsity pattern changes slowly enough, it is compressible along $T$. We use this idea to bound the error in CS on LSE and to show that if the sparsity pattern changes slowly enough, the CS-LSE error bound is much smaller than that of simple CS. 

We use the following definition of compressibility of the random process $\beta_t = \beta_t(x_t,y_{1:t})$.
\bd
We say that $\beta_t$ is {\bf compressible} if the maximum over $T$ of the average of $(\beta_t)_i^2$, conditioned on past observations, is smaller than the minimum average squared value of any currently nonzero component of $x_t$, i.e. if $\max_{i \in T} \E[ (\beta_t)_i^2 | y_{1:t-1}]  < \min_{i \in N_t} \E[(x_t)_i^2 ]$.
 This is a valid definition since $\min_{i \in N_t} \E[(x_t)_i^2 ] \le \min_{i \in \Delta} \E[(x_t)_i^2 ] = \min_{i \in \Delta} \E[(\beta_t)_i^2 ]$ for all choices $\Delta = \Delta(y_{1:t-1})$. 
\label{compdef0}
\ed
\vspace{-0.2in}

\begin{ass}[model, algorithm] 
Assume that
\ben
\item $y_t$, $x_t$ follow (\ref{obsmod}), (\ref{sysmod}); $w_t$, $\nu_t$ are independent of each other and over time; and $w_t$ has bounded support (e.g. truncated Gaussian) with cutoffs at $\pm \frac{\lambda_m \sigma_{obs}}{\max_i ||A_i||_1}$ in all dimensions.
\label{bndednoise}

\item  $N_{t-1} \subseteq N_t$ for all $t$ and $S_t := |N_t| \le S_{max}$.
\label{Ntincr}

\item  The number of false additions is bounded, i.e. $|T_t \setminus N_t| \le S_{fa}$ for all $t$. This implies that $|T_t| \le S_t + S_{fa} \le S_{max} + S_{fa}$. 
\label{falseadds_bnded}

\item  $\delta_{S_{max} + S_{fa}}< 1$. $\delta_S = \delta_S(A)$ is defined in \cite[eq. (1.3)]{dantzig}.
\label{delta_bnd}

\een
\label{model}
\end{ass}
\vspace{-0.1in}
Bounded measurement noise (Assumption \ref{model}.\ref{bndednoise}) is usually valid.
Assumption \ref{model}.\ref{falseadds_bnded} is observed in all our simulations, as long as the addition threshold $\alpha_a$ is large enough.
Assumption \ref{model}.\ref{delta_bnd} quantifies the required amount of incoherency of the measurement matrix w.r.t. the sparsity basis.
Consider Assumption \ref{model}.\ref{Ntincr}. While this assumption is not strictly true, it is observed (for medical image sequences) that it is approximately true: the set $N_t \setminus N_{t-1}$, and the total set $N_t \setminus N_{0}$, are both small. Also, if we relax the definition of support to denote any set containing all nonzero elements of $x_t$, then this is true.

Under the above assumptions, we can prove the following \cite{kfcsfullpap}:
\begin{theorem}
Assume that Assumption \ref{model} holds. Let $t_a=t_a(t)$ denote the last addition time before or at $t$.
\ben
\item
If $|\Delta|$ is small enough to ensure that
$(t-t_a+1) \sigma_{sys}^2 \ge \frac{\theta_{|T|,|\Delta|}^2}{(1-\delta_{|T|})^2} \lambda_{max}(\E[  (x_t)_\Delta (x_t)_\Delta ' | y_{1:t-1}]) + \frac{\sigma_{obs}^2}{1- \delta_{|T|}}
$,
then $\beta_t:=x_t - \xhat_{t,tmp}$ is compressible. $\theta_{S,S'}$ is defined in \cite[eq. (1.5)]{dantzig}.
\item The following bound on the CS-LSE error holds
\bea
\label{thm13givenT}
&& \E[|| x_t -\xhat_{t,CSLSE} ||_2^2 | y_{1:t-1}] \le \min_{1 \le S \le S_\infty} B_{CSLSE}(S) \nn \\
&& B_{CSLSE}(S) := C_2(S)S \sigma_{obs}^2 + C_3(S) \frac{(|T|+ |\Delta| - S)}{S} L0 \nn \\
&& L0 \defn \left\{ \begin{array}{ll}
\frac{\theta_{|T|,|\Delta|}^2}{(1- \delta_{|T|})^2} \E[ || (x_t)_\Delta||^2 |  y_{1:t-1}] + & \nn \\
(|T| + |\Delta| -S)\frac{\sigma_{obs}^2}{1- \delta_{|T|}}  & \ \ {\mbox{if}} \ \ S \ge |\Delta| \nn \\
(\frac{\theta_{|T|,|\Delta|}^2}{(1- \delta_{|T|})^2} + 1) \E[ || (x_t)_\Delta||^2 | y_{1:t-1}] +  &  \nn \\
|T|\frac{\sigma_{obs}^2}{1- \delta_{|T|}} & \ \ {\mbox{if}} \ \ S < |\Delta| 
\end{array} \right.
\eea
where $\xhat_{t,CSLSE}$ is the output of (\ref{ds}) with $\tty = y_t - A \xhat_{t,tmp}$ and $\xhat_{t,tmp} = (A_{T}'A_{T})^{-1} A_{T}' y_t$.
\een
\end{theorem}
Notice that $\E[  (x_t)_\Delta (x_t)_\Delta ' | y_{1:t-1}]$, and its trace, $\E[ || (x_t)_\Delta||^2 | y_{1:t-1}]$, can be computed by running a genie-aided KF. 

In \cite{kfcsfullpap}, we also derive a bound on the unconditional CS-LSE error, i.e. the error averaged over all values of the past observations, under slightly stronger assumptions.
We also discuss why the bound on the CS-LSE error is much smaller than that on simple CS error.

\Section{Convergence to Genie-Aided KF (or LS)}
\label{converge}
Consider the genie-aided KF, i.e. the KF which knows the true nonzero set, $N_t$, at each $t$. It is the linear MMSE estimator of $x_t$ from $y_1, \dots y_t$ if the nonzero sets, $N_t$'s, are known. It would be the MMSE estimator (i.e. it would be the best estimator among all possible estimators) if the observation noise were Gaussian instead of truncated Gaussian.
The genie-aided KF can be summarized as running (\ref{kfbig}) with $\hat{Q}_t =  \sigma_{sys}^2 I_{N_{t}}$.
In this section, we obtain conditions under which the KF-CS estimate converges to the genie-aided KF estimate in probability. As a corollary, we also get conditions for LS-CS to converge to genie-aided LS.


We begin by giving Lemma \ref{kfinitwrong} states that if the true nonzero set does not change after a certain time, and if eventually it is correctly detected, then KF-CS converges to GA-KF. This is followed by Lemmas \ref{nofalseadd_thm11bnd} and \ref{finitedelay} which prove that, if the addition threshold is high enough, the probability of false addition is zero and the probability of correct set detection approaches one with $t$. Combining these lemmas gives the final result.

\begin{lemma} \cite{kfcsfullpap}
Assume that there exists a $t_0$ s.t. $\forall \ t \ge t_0$, $T_t = N_t = N_*$ and assume that $\delta_{|N_*|} < 1$.
Consider KF-CS without the deletion step, i.e. with $\alpha_z = 0$, and with the step \ref{secondls} (KF-CS with final LS) replacing step \ref{secondkf} if $T_t \neq T_{t-1}$.
The difference in the KF-CS and GA-KF estimates, $d_t \defn  | \xhat_{t,GAKF} - \xhat_{t}|$, converges to zero in mean square and hence also in probability.
\label{kfinitwrong}
\end{lemma}


Assume that Assumption \ref{model} holds. The bounded support assumption on $w_t$ ensures that $|A_i'w_t| \le ||w_t||_\infty ||A_i||_1 \le  \lambda_m  \sigma_{obs}, \ \forall i$. With this, the theorems of \cite{dantzig} can be directly modified to hold with probability  one. This helps prove the following.
\begin{lemma}
Assume that (i) Assumption \ref{model} holds and that $\delta_{2 S_{max}}  + \delta_{3 S_{max}} < 1$ (stronger incoherency requirement than earlier); and (ii) in Algorithm \ref{kfcs}, we set $\alpha_a = B_1 \defn C_1^2 \lambda_m^2 S_{max} \sigma_{obs}^2$ ($C_1$ is defined in \cite[Thm. 1.1]{dantzig}).
Then, at each $t$, the following hold:
\bea
\label{thm11bnd}
|| x_t - \xhat_{t,tmp} - \betahat_t||^2 \le B_1 \defn C_1^2 \lambda_m^2 S_{max} \sigma_{obs}^2 \\
\label{nofalseadd}
\hat\Delta_t \subseteq N_t, \text{ and so } T_t \subseteq N_t, \text{ and so }  T_t \cup \Delta_{t+1} = N_{t+1}
\eea
\label{nofalseadd_thm11bnd}
\end{lemma}
\vspace{-0.2in}

{\em Proof: } 
We prove this result by induction.
At any $t$, when solving (\ref{ds}), $\xhat_{t,tmp,i} = 0, \ \forall i \in T_{t-1}^c$. The sparse vector to be estimated is $\beta_t \defn [ (x_t - \hat{x}_{t,tmp})_{T_{t-1}}, (x_t)_{\Delta_t}, 0_{N_t^c} ] $. 
First consider the base case, $t=1$. At $t=1$, $T_{t-1} = T_0 = \phi$ (empty) and so $\xhat_{1,tmp,i} = 0, \ \forall i$. Thus $\beta_1 = x_1$ with nonzero set $\Delta_1 = N_1$. Since $|N_1| \le S_{max}$ and since the observation noise, $w_t$, satisfies $|A_i'w_t| \le  \lambda_m  \sigma_{obs}$, we can apply Theorem 1.1 of \cite{dantzig} to get (\ref{thm11bnd}) to always hold at $t=1$. 

Also, for any $i \in N_1^c$, $x_{1,i} = 0$ and so  $\betahat_{1,i}^2 = (x_{1,i} - \betahat_{1,i})^2 \le || x_1 - \xhat_{1,tmp} - \betahat_1||^2 \le B_1$ (from (\ref{thm11bnd})). But $\alpha_a = B_1$. Thus, from (\ref{dsthresh}), $\hat\Delta_1 \subseteq N_1$. Thus $T_1 \defn T_{0} \cup \hat\Delta_1   \subseteq N_1$. But $N_1 \subseteq N_2$. Thus, $T_1 \subseteq N_2$. Since $\Delta_1 \defn N_1 \setminus T_0$, this implies that $T_1 \cup \Delta_2 = N_2$. Thus (\ref{nofalseadd}) also holds for $t=1$. This proves the base case. 

For the inductive step, assume that (\ref{nofalseadd}) and (\ref{thm11bnd}) hold for $t-1$.
Thus,  $T_{t-1} \cup \Delta_t = N_t$, which is the nonzero set for $\beta_t$. But $|N_t| \le S_{max}$.  Thus Theorem 1.1 of \cite{dantzig} can be applied to get (\ref{thm11bnd}) to hold for $t$.
Also, for any $i \in N_t^c$, $x_{t,i} = 0$ and so $\betahat_{t,i}^2 = (x_{t,i} - \betahat_{t,i})^2 \le || x_t - \xhat_{t,tmp} - \betahat_t||^2 \le B_1 = \alpha_a$.
Thus from (\ref{dsthresh}), $\hat\Delta_t \subseteq N_t$. Thus $T_t \defn T_{t-1} \cup \hat\Delta_t   \subseteq N_t \subseteq N_{t+1}$. Since $\Delta_{t+1} \defn N_{t+1} \setminus T_t$, this means that $T_t \cup \Delta_{t+1} = N_{t+1}$. Thus (\ref{nofalseadd}) holds for $t$. This proves the induction step and thus the result holds.

\begin{lemma}
Assume that (i) Assumption \ref{model} holds and that $\delta_{2 S_{max}}  + \delta_{3 S_{max}} < 1$; (ii) in Algorithm \ref{kfcs}, we set $\alpha_a = B_1 \defn C_1^2 \lambda_m^2 S_{max} \sigma_{obs}^2$; and (iii) all additions occur before a finite time, $t_{a,max}$, i.e. $N_t = N_{t_{a,max}}, \ \forall t \ge t_{a,max}$.
Let $N_* \defn N_{t_{a,max}}$. Then, 
$\lim_{t \tends \infty} Pr(T_{t+\tau} = N_{t+\tau} = N_*, \ \forall \tau \ge 0) = 1$
\label{finitedelay}
\end{lemma}

{\em Proof: }
Since (i) and (ii) hold, Lemma \ref{nofalseadd_thm11bnd} holds.
For any $i \in \Delta_t$, $\xhat_{t,tmp, i} = 0$. Thus, (\ref{thm11bnd}) implies that $(x_{t,i} - \betahat_{t,i})^2 \le B_1$ and so $|\betahat_{t,i}| \ge |x_{t,i}| - \sqrt{B_1}$.
Thus, if $|x_{t,i}| > \sqrt{B_1} + \sqrt{\alpha_a} = 2\sqrt{B_1}$, then $\betahat_{t,i}^2 > \alpha_a$, i.e. $i \in \hat\Delta_t$. In other words,
$
Pr(\{i \in \hat\Delta_t | x_{t,i}^2 > 4 B_1 \} ) = 1
$.
%
The same argument applies even if we consider all $i \in \Delta_t$. Thus, $Pr(\{\Delta_t \subseteq \hat\Delta_t \} | \{x_{t,i}^2 > 4 B_1 \ \forall i \in \Delta_t \}) = 1$. 

But from (\ref{nofalseadd}) and (\ref{dsthresh}), $\hat\Delta_t \subseteq \Delta_t$. Thus, if $x_{t,i}^2 > 4 B_1, \ \forall i \in \Delta_t$,   $\hat\Delta_t = \Delta_t$ and so $T_t \defn T_{t-1} \cup \hat\Delta_t =  N_t$. Thus, 
$
Pr(T_t = N_t | \{ x_{t,i}^2 > 4 B_1 \ \forall i \in \Delta_t \} ) = 1
$. 
Now,  $\forall t \ge t_{a,max}$, $N_t = N_*$. Thus for $t > t_{a,max}$,  $T_{t} = N_{*}$  implies that $\Delta_{t+1} = \phi$. This implies that $\hat\Delta_{t} = \phi$ and so $T_{t+1} = T_{t} = N_*$. Thus, $T_{t} = N_{*}$ implies that  $T_{t + k} = N_*, \ \forall k \ge 0$. Thus, for all $t > t_{a,max}$,
\bea
Pr(T_{t+\tau} =  N_* \ \forall \tau \ge 0  | \{x_{t,i}^2 > 4 B_1 \ \forall i \in \Delta_t\}) = 1   \ \ \ \ \ \ \ 
\label{detectall}
\eea
Now, $x_{t,i}^2 \sim \n(0, (t-t_i)\sigma_{sys}^2)$ where $t_i$ is the time at which element $i$ got added. Note that $t_i \le t_{a,max}$. Thus,
\bea
Pr( x_{t,i}^2 > 4 B_1 )  \ge  2Q (\sqrt\frac{4B_1}{(t-t_{a,max}) \sigma_{sys}^2}) \ \ \ \ \ \
\label{prxt}
\eea
where $Q$ is the Gaussian Q-function.
Combining (\ref{detectall}), (\ref{prxt}) and using the fact that the different $x_{t,i}$'s are independent, 
\bea
Pr(T_{t+\tau} = N_* \ \forall \tau \ge 0 )
\ge \left( 2Q (\sqrt\frac{4B_1}{(t-t_{a,max}) \sigma_{sys}^2}) \right)^{S_{max}} \ \ \ \ \
\eea
Thus for any $\eps > 0$,  $Pr(T_{t+\tau} =  N_*, \forall \tau \ge 0) \ge 1-\eps$ if $t \ge t_{a,max} + \tau_\eps$,
$
\tau_\eps \defn  \lceil  \frac{4 B_1}{ \sigma_{sys}^2 [Q^{-1}(\frac{ (1- \eps)^{1/S_{max}} }{2})]^2 } \rceil
$,
where $\lceil . \rceil$ is the greatest integer function. Thus the claim follows. $\blacksquare$


Combining Lemma \ref{finitedelay} with Lemma \ref{kfinitwrong} we get the final result.
%
%
%
\begin{theorem}
Assume that (i) Assumption \ref{model} holds and that $\delta_{2 S_{max}}  + \delta_{3 S_{max}} < 1$; (ii) in Algorithm \ref{kfcs}, we set $\alpha_a = B_1 \defn C_1^2 \lambda_m^2 S_{max} \sigma_{obs}^2$; and (iii) all additions occur before a finite time, $t_{a,max}$, i.e. $N_t = N_{t_{a,max}}, \ \forall t \ge t_{a,max}$.
Consider KF-CS without the deletion step, i.e. with $\alpha_z = 0$, and with the step \ref{secondls} (KF-CS with final LS) replacing step \ref{secondkf} if $T_t \neq T_{t-1}$.
Then,
$d_t \defn \xhat_{t,GAKF} - \xhat_{t}$ converges to zero in probability, i.e. the KF-CS estimate converges to the  Genie-Aided KF estimate in probability, as $t \tends \infty$.
\\
Also, the LS-CS estimate converges to the  Genie-Aided LS estimate, in probability, as $t \tends \infty$ (this follows directly from Lemma \ref{finitedelay}).
\label{thm1}
\end{theorem}
\vspace{-0.05in}
The assumption $\delta_{2 S_{max}}  + \delta_{3 S_{max}} < 1$ is just stronger incoherency requirement on $A$ than Assumption \ref{model}.
The assumption of all additions occurring before a finite time is a valid one for problems where the system is initially in its transient state (nonstationary), but later stabilizes to a stationary state. Alternatively, the above theorem can be applied to claim that the KF-CS (or LS-CS) estimate stabilizes to within a small error the GA-KF (or GA-LS) estimate,  if additions occur slowly enough, i.e. if the delay between two addition times is long enough to allow it to stabilize \cite{kfcsfullpap}.

%
%

\Section{Simulation Results}
\label{sims}

Lemma \ref{nofalseadd_thm11bnd} says that if the addition threshold was set high enough ($\alpha_a = B_1$ where $B_1$ is the CS error upper bound), then there would be no false additions. But if we set the addition threshold very high, then for the initial time instants, the KF-CS estimation error would be large and it will do worse than simple CS. Thus, in practice, we set $\alpha_a$ lower, but we implement the false addition detection and removal scheme described in Sec. \ref{kfcswithdel}.
We evaluated its performance using the following set of simulations. We simulated a time sequence of sparse $m$=256 length signals, $z_t = x_t$ which follow (\ref{sysmod}) with $\sigma_{sys}^2=1$ and nonzero sets, $N_{t-1} \subseteq N_t, \ \forall t$ satisfying $N_t = N_1, \forall t < 10$, $N_t = N_{10}, \forall 10 \le t < 20$,  $N_t = N_{20}, \forall 20 \le t < 30$, $N_t = N_{30}, \forall 30 \le t < 100$ and $|N_1| = 8$,  $|N_{10}| = 12$, $|N_{20}| = 16$, $|N_{30}| = 20$. Thus $S_{max}=20$. The set $N_1$ and all the additions were generated uniformly at random from the remaining elements out of $[1:m]$. The measurement matrix, $A = H$ was simulated as in \cite{dantzig} by generating $n \times m$ i.i.d. Gaussian entries (with $n=72$) and normalizing each column of the resulting matrix. The observation noise variance was $\sigma_{obs}^2 = ( (1/3)\sqrt{16/n})^2$ (this is taken from \cite{dantzig}) and we simulated Gaussian noise (not truncated).

We implemented KF-CS with $\lambda_m = \sqrt{2 \log_2 m}=4$, $\alpha_a = 9 \sigma_{obs}^2$, $\alpha_{fe} = 2n$, $\alpha_{z} = \sigma_{obs}^2$, $k = 5$, $k' = 3$. Since the observation noise was not truncated, occasionally the addition step resulted in a very large number of false additions, which made $A_{T_t}'A_{T_t}$ singular (or almost singular) resulting in large errors at all future $t$. To prevent this, we set a maximum value for the number of allowed additions: we allowed at most $(1.25n/\log_2 m)$ largest magnitude coefficient estimates larger than $\alpha_a$ to be added. Also, typically an addition took 2-3 time instants to get detected. Thus we set $\sigma_{init}^2 = 3 \sigma_{sys}^2$ ($\sigma_{init}^2$ is used instead of $\sigma_{sys}^2$ the first time a new coefficient gets added).
We simulated the above system 100 times and compared the MSE of KF-CS with that of GA-KF and of simple CS (followed by thresholding and least squares estimation as in Gauss-Dantzig \cite{dantzig}). 

In a second set of simulations, shown in Fig. \ref{sims_del}, we started with $S_1=8$ and for $10 \le t \le 50$, we added 2 new elements every 5 time units. Thus $S_{max}=26 = S_t, \forall \ t \ge 50$. Note $26 > n/3 = 24$, i.e. $\delta_{3S_{max}}$ cannot be smaller than 1.


\Section{Discussion and Ongoing Work}
\label{conclusions}
In this work, we introduced Least Squares CS and analyzed why CS on the LS error in the observation will have lower error than CS on the raw observations (simple CS), when sparsity patterns change slowly enough. 
We also showed that if all additions occur before a finite time, if the addition threshold is set high enough, if UUP holds for $S_{max}$, and if the noise has bounded support, KF-CS (or LS-CS), converge to the genie-aided KF  (or LS) in probability. In ongoing work, we are working on relaxing the first three assumptions used in the above result. We are also working on developing KF-CS for real-time dynamic MR imaging \cite{kfcsmri}.

\vspace{-0.1in}

\bibliographystyle{IEEEbib}
\bibliography{kfcspap_camready}

\end{document}